# Ambient-Pressure X-ray Photoelectron Spectroscopy through Electron Transparent Graphene Membranes

*Jürgen Kraus[1], Robert Reichelt[1], Sebastian Günther[1], Luca Gregoratti[2], Matteo Amati[2], Maya Kiskinova[2],*

*Alexander Yulaev[3,5], Ivan Vlassiouk[4] and Andrei Kolmakov[5]\**

[1]Technische Universität München D-85748 Garching, Germany

[2] Sincrotrone Trieste 34012 Trieste, Italy

[3]Department of Materials Science and Engineering, University of Maryland, College Park, MD 20742, USA

[4] Oak Ridge National Laboratory, Oak Ridge, TN 37831, USA

[5] Center for Nanoscale Science and Technology, NIST, Gaithersburg, MD 20899, USA

**Abstract**

Photoelectron spectroscopy (PES) and microscopy are highly demanded for exploring morphologically complex solid-gas and solid-liquid interfaces under realistic conditions, but the very small electron mean free path inside the dense media imposes serious experimental challenges. Currently, near ambient pressure PES is conducted using sophisticated and expensive electron energy analyzers coupled with differentially pumped electron lenses. An alternative economical approach proposed in this report uses ultrathin graphene membranes to isolate the ambient sample environment from the PES detection system. We demonstrate that the graphene membrane separating windows are both mechanically robust and sufficiently transparent for electrons in a wide energy range to allow PES of liquid and gaseous water. The reported proof-of-principle experiments also open a principal possibility to probe vacuum-incompatible toxic or reactive samples enclosed inside the hermetic environmental cells.

*Corresponding author E-mail: andrei.kolmakov@nist.gov

Ambient pressure *in situ* photoelectron spectroscopy and imaging of submicrometer structured matter and interfaces under operating conditions are characterization tools that are highly demanded in many research areas, including catalysis[1], fuel cells[2], batteries[3] and bio-medical devices[4]. However, such experiments are difficult to perform, since the chemical and spatial information carried by the emitted electrons is lost upon first inelastic collision in the dense medium. Therefore, the emitted electrons have to be collected at a distance from the specimen comparable to the electron inelastic mean free path (IMFP), which for electron kinetic energies between $10^2$ eV and $10^3$ eV is about 1 nm for liquids and 1 µm for ambient pressure gases. Historically, this experimental challenge of bridging the so-called "pressure gap" in electron spectroscopy has been partially resolved by employing liquid beams (jets [5], droplets on the fly [6]), wetted discs and rods [7] or sophisticated engineering of the differentially pumped electron energy analyzers[8] as well as pulsed gas delivery[9] in combination with ultra-bright synchrotron radiation sources. These approaches, reviewed elsewhere[10], have allowed PES measurements at pressures of a few hundred Pa of reactive gases and liquids. To respond to the rapidly growing research demand for near-ambient pressure PES (APPES), affordable laboratory-based APPES instruments have been developed as well recently[11].

Using ultra-thin membranes sufficiently transparent for electron transmission can be an alternative approach to separating the high pressure sample environment from the ultra-high vacuum (UHV) in the analyzer. In spite of its long history (see [12] and references therein), this approach has suffered from the lack of membranes that are sufficiently thin and yet mechanically robust enough to sustain the pressure differential. The membrane thickness constraints however can be fairly relaxed if hard X-rays are used for PES, since photoelectrons in this case have high kinetic energies and therefore longer IMFPs[13].

The recent progress in large-scale fabrication and handling protocols of novel two-dimensional

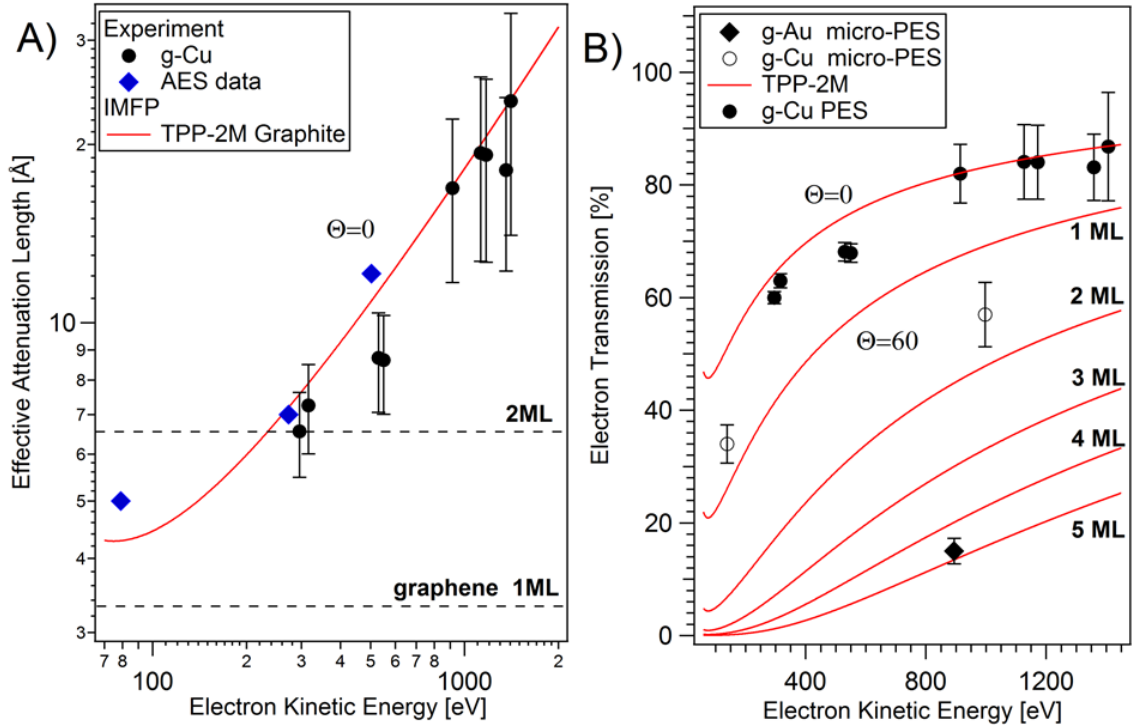

**Figure 1**. A) Solid line: predicted inelastic mean free path in graphite as calculated using the optical data [23] and TPP-2M formula with the set of parameters from Ref. 18. Experimental data points (filled circles) represent the measured effective attenuation length (EAL) for graphene monolayer on Cu using Cu 2p, Cu 3s, Cu 3p and Cu LMM electrons irradiated with Mg K$\alpha$ and Al K$\alpha$ photons; the electron emission angle $\theta = 0°$. In addition, EAL data (diamonds) from Ref. 18 are included. Horizontal dashed lines indicate the thicknesses of single and bilayer graphene. B) Experimental data for electron transmission through a single graphene layer in a laboratory PES setup ($\theta = 0°$ filled circles) and a µ-PES setup ($\theta = 60°$; open circles). The solid lines show the expected electron transparency ($I/I_0$) of graphene with the indicated layer thickness as calculated using the TPP-2M IMFP formula and corresponding emission angle. Filled diamond corresponds to wet transferred graphene with a nominal thickness of 4ML and measured in the µ-probe PES setup. The data show that the as-grown g-Cu samples were covered by a 1to 2 ML graphene layer, while the wet processed and transferred graphene of the g-Au sample had an effective thickness of about 5 ML (including contaminants; see text for details). The systematic errors in attenuation and transmission measurements shown in the panels A and B as error bars exceed the statistical errors and are related to reproducibility of the X-ray photon flux and electron detection efficiency after each measurement (see Methods Section for details)

materials such as graphene, graphene oxide, boron nitride, *etc.*[13], have ignited intensive research studies of their exotic physical and chemical properties as well as applications[14]. The properties of these materials have revived the idea of electron transmission and spectroscopy through membranes that are reasonably transparent to photoelectrons with relatively low kinetic energy (less than 1 keV)[15]. The quantification of the transparency of the graphene membranes is directly related to the standard surface science problem of the attenuation of the substrate signal by thin overlayer film. The experimentally accessible parameter such as electron attenuation lengths (EAL) is commonly used in this case[16]. It is notable that EALs for these low-Z single (or two) monolayer (ML) thick materials exceeds or are comparable to their thicknesses at energy as low as 300 eV. Semi-empirical calculations indicate (see Figure 1) that the electron transparency of single layer graphene could be greater than 50 % for electron energies greater than about 300 eV [17]; only elastic scattering will limit the transmissivity of a suspended single monolayer graphene membrane. The few available electron attenuation measurements[18] (including data reported in this work) support these predictions (see data points in Fig. 1). These results, in conjunction with the reported mechanical stiffness and gas impermeability of graphene membranes [19], have opened an alternative opportunity to probe a very broad classes of materials and interfaces. Reactive, toxic or radioactive materials in any state of aggregation can be tested using powerful electron spectroscopy and microscopy tools *i.e.* under experimental conditions which were not achievable before[20].

Here we report proof-of-principle results demonstrating that, using suspended graphene (or in principle any other molecularly impenetrable 2D material) as a separating membrane between vacuum and a liquid or dense gaseous medium, PES spectra of sufficient quality and low-electron-energy scanning electron microscopy (SEM) images of objects immersed in a liquid water environment can be recorded . In this report, we tested and compared two designs of graphene-based single-use windows for *in situ* PES and characterized the achievable transparency for photoelectrons with different kinetic energies less than $10^3$ eV. Due to the small size of the membrane covered orifices, the PES experiments for probing the matter behind the suspended graphene-based membranes were performed using a focused X-ray beam (μ-probe PES or μ-PES)[21]. The attenuation lengths of the photoelectrons at several kinetic energies were measured for graphene-covered macroscopic calibration samples, using a standard PES. The feasibility of the performing PES through the graphene windows has been demonstrated by probing liquid water with μ-PES. Limiting factors such as water radiolysis at high irradiation doses are discussed.

Monolayer graphene was grown on 25 μm thick Cu foils (g-Cu) at a temperature of ≈1000 °C in a CH$_4$/H$_2$ reactive atmosphere in the 30 Pa to 5000 Pa range following a protocol described in Ref [22]. For the fabrication of more robust graphene windows for liquid cells (so called environmental cells or E-cells) membranes with an average thickness of 4 ML were grown by chemical vapor deposition (CVD) deposition Ni/Si (Ni films deposited on Si wafers) and wet transferred on to Au pre-covered support samples (g-Au) which contained orifices with width of a few micrometers. The design and tests of the E-cell as well as of the graphene-transfer protocols are described elsewhere[20c]. All samples were transported through air to the laboratory PES system or to the scanning PES microscope at the synchrotron facility ELETTRA, where E-cell assembly, filling with water, sealing and vacuum tightness tests were performed prior to the μ-PES measurements. The PES spectra from the graphene-covered and adjacent bare areas of the Au substrate were measured for comparison (Fig. 2). In addition, to demonstrate the quality of the spectroscopy through graphene, we recorded spectra of suspended membranes and compared them with those obtained from the same membranes after Au had been deposited on their back side. Finally, as a proof of principle, dynamical changes of the O1s photoelectron spectrum from a water droplet were obtained from electrons transmitted through the graphene membrane of a vacuum-tight E-cell.

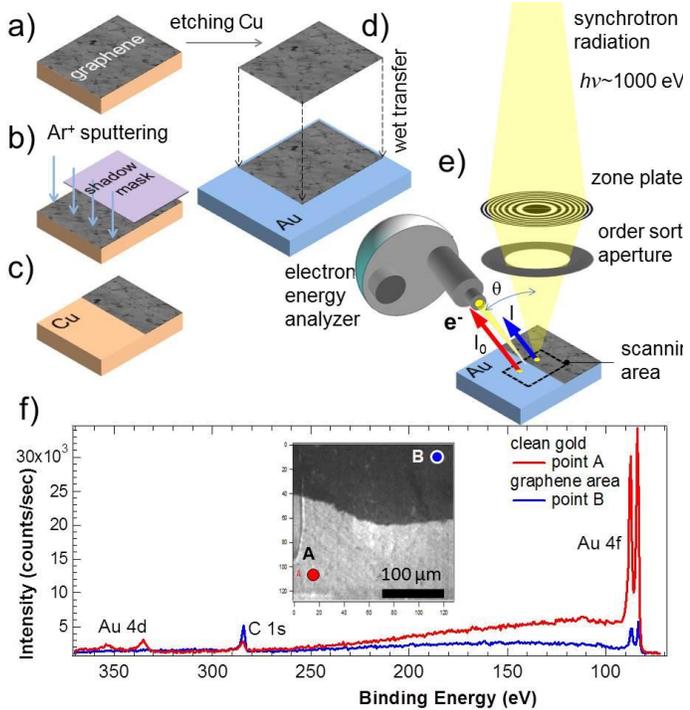

**Figure 2**. Sample preparation and experimental setup: a) as grown G layer on a Cu foil (g-Cu-1); b) Ar$^+$ sputtering of half of the sample using the shadow mask; c) Resultant pristine and graphene-covered Cu substrate for comparative XPS analysis; d) Alternative sample design using PMMA-based transfer of graphene on to a Au surface; e) Experimental setup of scanning photoelectron microscope at ELETTRA; f) (inset) 256 μm x 256 μm XPS map of the half graphene-covered Au sample taken by acquiring Au 4f photoelectrons; Comparison of the XPS survey spectra acquired from points A and B in the inset. Significant attenuation of Au 4d, 4f peaks by the 4 ML thick graphene layer can be noticed.

**Results and discussion**

*Attenuation tests for single and multilayer g-samples*

The measurements of overlayer attenuation of the photoelectron signal were performed on as-grown graphene model samples and compared with values from the IMFP predictive formula for graphite by Tanuma, Powell and Penn[23] (TPP-2M) and with available experimental data from Auger electron spectroscopy (AES)[18]. For this purpose Cu 2p, Cu 3s, Cu 3p photoelectron and Cu LVV Auger peak intensities were recorded comparatively before ($I$) and after ($I_0$) removal of the monolayer graphene by prolonged Ar$^+$ ion etching using a laboratory XPS system. Assuming the thickness of the graphene layer as $d_G$ = 3.35 Å, the electron effective attenuation length $\lambda_{EAL}$ in monolayer graphene was estimated from measured $I/I_0$ intensity ratios using standard overlayer-film attenuation formula[17]:

$$\lambda_{EAL} = \frac{d_G}{\ln(I_0/I)\cos\theta}$$

(for the laboratory attenuation tests, the electron emission angle for the detected electrons was θ = 0°). These practical electron attenuation length (EAL) values will be used later to evaluate the thickness of the suspended and wet-transferred graphene membranes. The experimental values for the EAL are displayed in Fig. 1A together with the IMFP for graphite calculated applying the predictive TPP-2M equation (solid curve). In addition, the data from reference[18] are plotted in the graph. One can observe that experimental data agree reasonably well with the values from the TTP-2M predictive formula using the set of parameters for graphite[18] and with IMFP calculations based on optical data[23]. The systematic deviation of the EAL data to lower values compared to the IMFPs is due to elastic scattering of electrons, which is taken into account in the $\lambda_{EAL}$ values. It has been shown[16] that, deviations of practical EALs from the corresponding IMFP values can be as high as 35 % depending on energy, sample thickness and emission angle.

We can now complement the electron attenuation data with electron transparency ($I/I_0$) data for the as-grown (g-Cu) and the transferred graphene layers (g-Au) that were characterized in the μ-PES setup using the spectrometer with an emission angle θ = 60°. The experimental μ-PES data points on graphene transparency displayed in Fig. 1b indicate that significant intensity of photoelectrons can be obtained from the samples placed behind graphene membranes just a few ML thick. Figures 2e and 3a show the experimental setup and geometry of the µ-PES attenuation measurements, the acquired spectra, and chemical maps of the two classes of samples: (i) transferred graphene on to an Au-coated substrate (g-Au) and (ii) as-grown graphene on Cu (g-Cu).

Figure 2f displays the measured signal attenuation of the Au 4f photoelectrons originating from the g-Au sample, which was half-covered by a transferred 4 ML thick graphene. Here, the Au 4f chemical map identifies the edge of the wet chemically transferred graphene film, i.e. it images the border between the graphene-covered and the bare Au surface. The spectra acquired from the graphene-covered and the graphene-free areas reflect the considerable damping of the substrate signal by the wet-transferred graphene layer. We determined the effective signal attenuation for the Au 4f photoelectrons of 85 % (at kinetic electron energy of 894 eV). Extracting the precise $\lambda_{EAL}$ values for Au sample partially covered with wet-transferred graphene using the aforementioned algorithms would be unreasonable since the wet transfer procedure unavoidably leads to the surface contamination of the graphene overlayer with no *a priori* known thickness. However, it is interesting to evaluate the degree of graphene contamination for this standard transfer protocol relating to the EAL expected from TTP-2M. Electron intensity attenuation by a graphene layer with a nominal thickness of 4ML was measured comparing Au 4f intensities collected from adjacent g-covered versus uncovered areas (Fig. 2f). Taking $\lambda_{EAL}$=16.7 Å for $E_k$=893 eV Au 4f electrons and the measured ratio of $I/I_0$=0.15 (see Fig. 1b) one can evaluate the effective thickness of the transferred graphene as ≈ 5 ML which corresponds to at least one additional monolayer of contaminants. The possible sources of the contamination are the remnant nanoscopic patches of the poly(methyl methacrylate) (PMMA) protection layer and solvents residue, which indeed can be observed, e.g. in the XPS data as Si 2p traces with an intensity corresponding to 0.4 ML.

More reliable data can be obtained from the g-Cu sample whose one half was sputter cleaned, while the other half was protected underneath by a shadowing mask (see Fig. 2a-c), generating a border region between the graphene-covered and graphene-free Cu. The acquired Cu 2p map of the border region

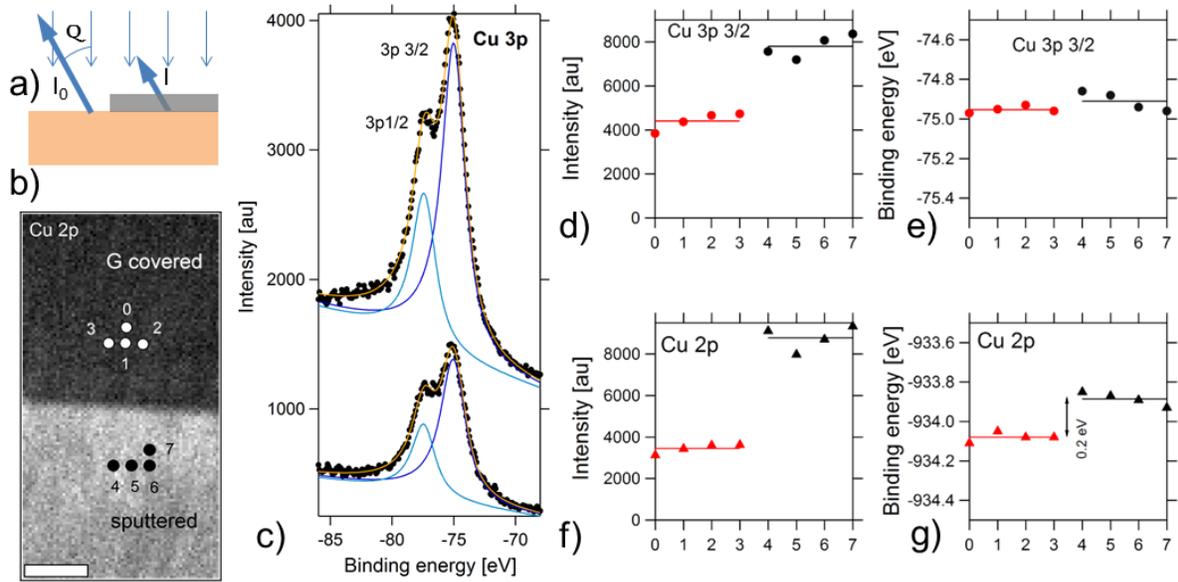

**Figure 3** a) The geometry of the attenuation tests on g-Cu samples; b) Cu 2p chemical map near the border of the g-covered and sputtered regions of g-Cu sample. The scale bar corresponds to 100 µm. The numbered points represent the locations; c) typical collected Cu 3p doublet spectra acquired from g-covered (bottom) and graphene-free (top) Cu regions with example fits of these spectra; d) corresponding Cu $3p_{3/2}$ peak intensities obtained from g-covered (points 0-3) and g-free (points 4-7); e) Cu $3p_{3/2}$ peak position as a function of the location; f) and g) the similar dependences as in d) and e) recorded for Cu $2p_{3/2}$ spectra (with lower kinetic energies than for the Cu 3p spectra) at the same locations where Cu 3p spectra were collected is shown in Fig 3 b. Several spectra were recorded at different locations on the graphene-free (black dots) and the graphene-covered parts of the Cu foil (white dots). Figure 3c, shows the Cu 3p doublet spectra acquired on the graphene-covered and the graphene-free Cu region. The lower intensity for the bottom spectrum is due to attenuation of the emitted Cu 3p photoelectrons by the graphene layer. The fitted intensities for the Cu $2p_{3/2}$ and Cu $3p_{3/2}$ peaks (acquired at the different indicated locations) displayed in Fig. 3d and 3e, correspond to a signal attenuation of 66 % for Cu 2p (electron kinetic energy 138 eV) and 43 % for the Cu 3p (electron kinetic energy 997 eV). Both data points are plotted as transmission % in Fig. 1b for comparison with the already mentioned analytic electron transparency curves (TPP-2M equation, assuming $d_G = 3.35$ Å, $\theta = 60°$). As can be seen, the observed signal attenuation in µ-PES experiments corroborates with a graphene layer thickness ranging between 1 and 2 ML. The observed deviation from the predictive equation most likely stems from the enhanced contribution of the elastic scattering for high values of emission angles and the fact that the thickness of the graphene in the g-Cu sample was not homogeneous on the micro-scale. The latter was confirmed by Raman spectra, which revealed the presence of the bilayer patches and defects in the graphene as indicated by the existence of a D-band and the large full width at half maximum (FWHM) of the G and 2D bands (Fig. S1 in Supporting material). The deviations could also be partially due to insufficient data averaging, since the spectra acquired in the µ-PES contain substantial intensity variation due to the local topography of the sample surface[24].

In addition, comparing Cu 2p spectra recorded at graphene-free and graphene-covered areas we observed a small (≈ 0.2 eV) but consistent binding energy shift in the Cu $2p_{3/2}$ peaks positions (Fig. 3g). Although the interfacial interaction between graphene and Cu transition metal is traditionally considered to be weak, restructuring of the Cu support underneath the CVD grown graphene was recently reported [25, 26]. Thus, we speculate that the observed energy shift might be related to a modified surface structure of Cu underneath the grown graphene layer, which apparently was altered by the graphene removal upon $Ar^+$ sputtering. This explanation is supported by almost vanishing of the aforementioned energy shift in the Cu 3p spectra (Fig. 3e). The latter is due to the fact that the escape depth of the 1000 eV 3p Cu electrons is greater than 1.4 nm, which makes them insensitive to the graphene induced surface reconstruction.

*Electron transparency of free suspended graphene membranes under soft X-ray excitation*

To investigate the electron transparency of the as-prepared suspended graphene membranes for ambient pressure PES, we used model g-Cu samples whose back-side was locally etched under controlled conditions producing graphene covered micro-holes as outlined in the Methods section and the Supporting Material. As is shown in Fig. 4a, the etching of Cu foil was monitored in real time by optical microscopy and interrupted after the first micro-holes in the Cu support were observed. Since the protecting PMMA layer was not used during this etching technique the membranes should be ultimately clean and highly electron transparent. For the sample depicted in the SEM image in Fig. 4c, three suspended membranes: I, II, and III were formed after the etching procedure. The scanning transmission X-ray microscopy (STXM) image next to the SEM image was obtained by recording the transmitted light, which proves that indeed through micro-holes were etched in Cu support. As it is shown in the supporting

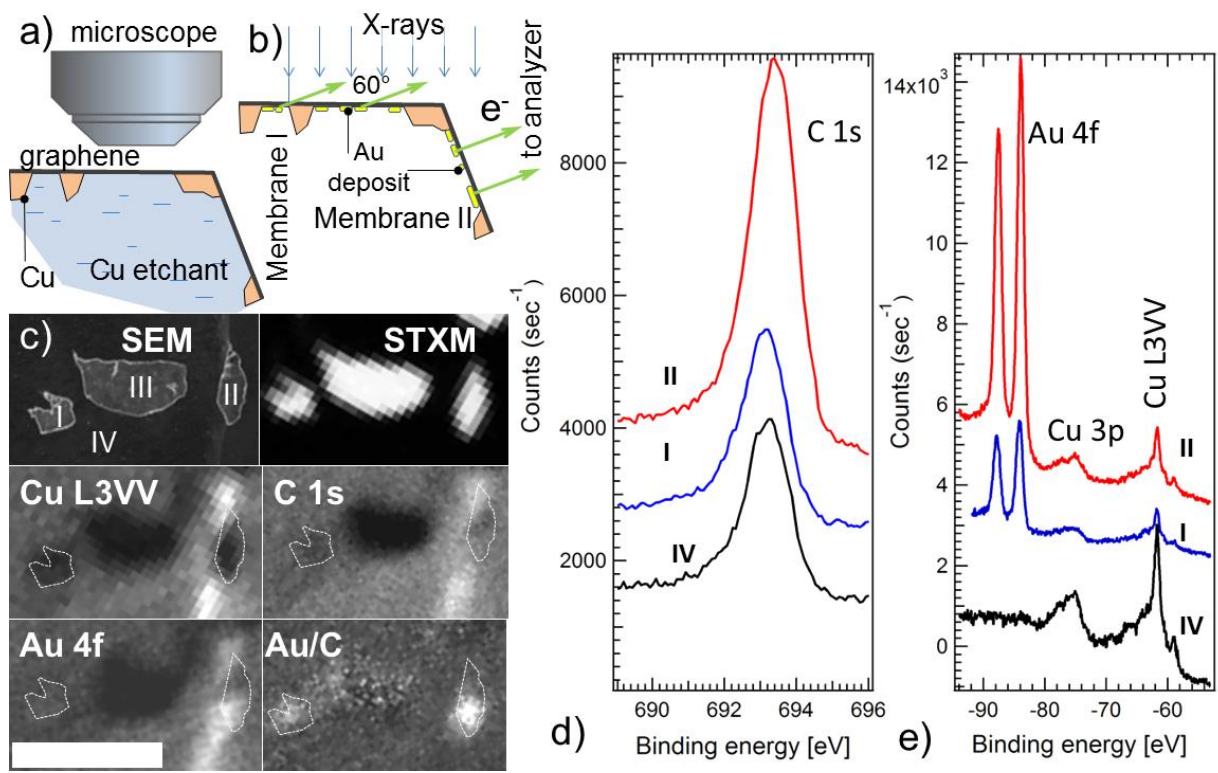

**Figure 4** a) Optical microscopy setup for real time observation of g-Cu foil etching process; b) After etching the sample has three suspended graphene membranes I, II and III, where membrane II was located at a Cu foil facet which was differently inclined with respect to the X-ray beam and electron energy analyzer; c) SEM, STXM images and the set of XPS chemical maps showing the presence of the gold deposits on the backside of suspended graphene membreanes I and II. The scale bar corresponds to ten micrometers; d) C 1s spectra taken at the membranes I (blue), II (red) and at the position IV (graphene-covered Cu area close to membrane I); e) Correspnding Au 4f spectra (the indexing is the same as in d).

material the membrane II was formed along a grain boundary of the Cu support foil which was highly bent (by 60°) with respect to the main part of Cu foil. As a result, membranes I and II conveniently have a different geometric inclination with respect to the irradiating X-ray beam and the photoelectron analyzer, which is depicted in Fig. 4b. Due to the different local geometry of the membranes I and II with respect to the electron analyzer, these two membranes have different attenuation ability *i.e.* membrane II should be more transparent for electrons, when compared with membrane I. In order to employ this geometrical factor on the membranes transparency evaluation, a gold layer was deposited simultaneously on the backside of the two graphene membranes (Fig. 4b) and the sample was imaged before and after the Au deposition using the SPEM apparatus. The figure 4c contains a series of chemical maps of the same area recorded in Cu L3VV, C 1s and Au 4f photoelectrons after a layer of Au was deposited on the backside of the membrane. Since the foil substrate was bent (60°) along the line, which is seen as a bright rim spreading in the vertical direction in Cu L3VV, Au 4f and the C 1s maps displayed in Fig 4c, the photoemission signal intensity of the rim in these maps can be attributed to the surface topography as described in details in the supporting material. The bottom left panel of Fig. 4c shows the Au 4f map of the area after Au deposition. To eliminate a strong topography contribution, the Au 4f image was divided by the subsequently recorded C 1s image. The result is shown in the lower right panel of Fig. 4c. The enhanced Au 4f / C 1s grey level intensity on membrane I and II is clearly visible as expected due to the Au 4f photoelectron emission from the backside of the membranes. The higher contrast from membrane II is evident as well. This experiment demonstrates a principal possibility to collect sufficient quality XPS signal through the suspended membrane after wet etching and rinsing steps as well as an importance of the illumination and collection geometry of the experiment. The enhanced brightness of the membrane II compared to I is due to: (i) increased amount (by a factor of *$(cos\Theta)^{-1}$; $\Theta=60°$*) of irradiated Au atoms due to the X-ray grazing incidence angle; (ii) photoelectron emission angle of about *0°* with respect to the surface normal, which leads to a larger signal contribution from photoemission of deeper layers and (iii) only to a minor extent, due to a changed transparency of the membranes in the different geometries.

More quantitative and detailed analysis as well as an estimation of the detection limit of photoemission experiments through a monolayer thick graphene membrane can be achieved via comparing the intensities and peaks positions of appropriate XPS spectra (see section 4 of supporting materials for details). Figures 4d, e display C 1s and Au 4f spectra acquired from the planar membrane I and inclined membrane II, respectively. In addition, the corresponding spectra from the regular graphene-covered Cu substrate (IV) are presented for comparison. The intensity comparison of Au 4f peaks taken at the membranes I and II comply in general with the aforementioned geometric consideration. The quantitative analysis of Au/C XPS intensities can be used for detection limit determination for soft X-ray photoemission spectroscopy through a monolayer-thick graphene membrane. It is known that Au follows Volmer-Weber growth on graphite in accordance with the literature[27]. Assuming the validity of this growth mode for Au on graphene, the lack of any noticeable chemical Au 4f shift in our experiments indicates that the formed Au clusters are significantly larger than 10 nm in diameter with an average island thickness larger than 2.5 nm. This assumption greatly simplifies the calculations of the detection limit. The minimum detectable amount of gold by Au 4f photoemission through a suspended monolayer graphene membrane in the geometry I and II was estimated to be lower than 1 % and 0.25 % of a monolayer respectively. This high sensitivity clearly shows the potential of graphene-based membranes for sensitive probing the matter behind the membrane.

*Ambient pressure PES through g-membranes*

To demonstrate the capability of ambient pressure PES through a graphene membrane, a ≈ 10 μL droplet of ultrapure water was placed on the membrane backside of an E-cell which was sealed in air to isolate the droplet from UHV of the SPEM. Figures 5a and 5b show the chemical maps of the g-window and its surroundings acquired by collecting O 1s photoelectrons. Topographical features dominate the contrast in each image. Figure 5b depicts the area of Fig. 5a after a prolonged exposure of point B inside the membrane window to the focused X-ray beam where the μ-PES data acquisition took place. Figure 5c represents a set of O 1s spectra collected at point B sequentially starting from the second one from the bottom. The very bottom spectrum was taken at point A (Fig 5a) at about 20 μm away from the membrane while the rest of the spectra were recorded at point B within the membrane window (Fig. 5b). The deconvolution of the O 1s spectrum was performed using three components, which are assigned to OH, $H_2O$ (L) and $H_2O$ (V), where *L* and *V* stand for liquid and vapor respectively. These assignments are based on the relative positions and chemical shifts for surface hydroxyl groups, adsorbed molecular water and vapor measured with APPES of hydrated oxides[28] and micro jets [5c]. In our experiments, the existence of molecular water and hydroxyl adsorbed species beyond the orifice can be attributed to a water monolayer that was trapped at the graphene-substrate interface during the wet transfer protocol and /or diffused from the orifice[29]. With respect to the O 1s spectrum acquired from the graphene covered support (point A), the first O 1s spectrum recorded from the membrane (point B) is significantly broadened and enhanced in peak intensity. We assign these differences to the presence of liquid water behind the membrane which causes an increased photoelectron emission from deeper layers of the liquid water. The set of spectra in the Figure 5b compiles O 1s spectra which were sequentially recorded requiring an acquisition time of ≈ 30 s per spectrum. There is a noticeable redistribution in the O 1s peak components as a function of the exposure time, which is quantitatively depicted in the bottom panel of the Fig 5c. As can be seen, after about two minutes of exposure to the focused X-ray beam, there was a sudden appearance and gradual growth of a new O 1s component at the cost of the one belonging to the liquid water $H_2O$ (L) component. Since this new O 1s component can be attributed to O 1s emission from water vapor $H_2O$ (G), the observed process can be attributed to X-ray beam-induced micro-bubble formation underneath the membrane, which also accounts for the bright spot visible in the SPEM image of Fig. 5b. Electron beam-induced micro-bubble formation and growth in liquids is regularly reported in the literature (see recent ref.[30] and references therein). Very similar dynamics was observed during scanning electron microscopy of the liquid water encapsulated by a graphene membrane[20c], where we found that a 20 keV electron beam irradiation of liquid water under graphene membrane does not lead to immediate bubble formation but requires the accumulation of a critical energy dose of ≈ $10^8$ eV·nm$^{-2}$ to $10^9$ eV·nm$^{-2}$ to nucleate the first bubble. These values match the ones estimated from µ-PES experiment. To support this interpretation further, Fig. 5d depicts four snapshots of a SEM video sequence recorded from water below a sealing graphene membrane, where the time delay between each displayed image amounts to 10 s to 15 s. The reduced density of the gaseous medium lowers secondary electron yield in the SEM images; therefore, the darker round area corresponds to a gas bubble formed at the interface between water and graphene. Under continuous electron beam rastering, the bubble increases in size until occupation of the entire diameter of the g-window (see the video in the Supporting Material).

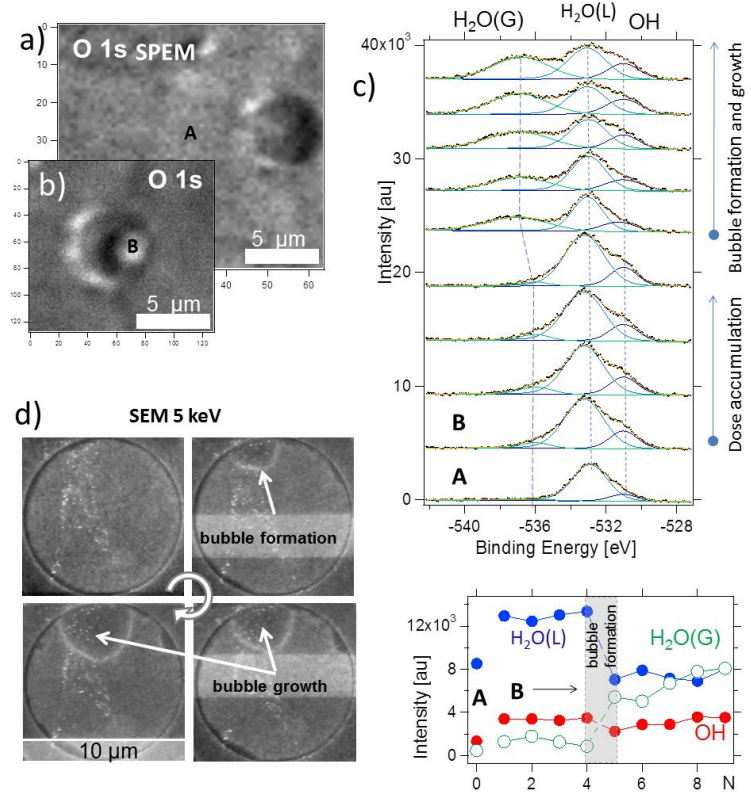

**Figure 5.** a) O1s XPS chemical maps of the graphene covered orifice before and after (b) local spectra acquisition; c) time evolution of the O1s spectra taken from the location B on the membrane; spectrum A was recorded on the graphene covered support outside the membrane (point A in Fig.5a) and is displayed for comparison; Bottom panel represent the intensity of the deconvoluted O 1s spectra as a function of time; d) SEM images of the beam-induced gas bubble nucleation and growth.

Following the enlightenments reported in the previous studies of liquid water[20c, 30-31], frozen hydrated samples [32], the bubble formation is primarily a result of water radiolysis by intense ionizing radiation and heating is a secondary effect . In our case, the X-ray energy dissipates in liquid water via creation of a variety of ionized and excited molecular species in the interaction volume. Most of the ions and radicals recombine rapidly, but a few chemically reactive products such as molecular hydrogen ($H_2$), hydrogen peroxide ($H_2O_2$) and hydroxyl radicals (·OH) accumulate [31a] and eventually segregate as a separate phase towards the hydrophobic graphene membrane. As a result, hydrogen, peroxide and water vapor-containing bubbles are formed under the graphene membrane leading to a potential pressure buildup inside the E-cell and chemical etching of the membrane. Therefore, from the practical point of view the observed X-ray induced dynamic processes in liquid water are unwanted, since they obstruct interfacial processes and eventually destroy the membrane. The natural solution of this experimental challenge will be the optimization of the membrane design and a compromise during adjustment of the lateral resolution, photon flux, electron collection dwell time, pass energy of the analyzer, etc. to the specific needs of the experiment. Alternatively, for a large class of experiments high spatial resolution is not required. To extend our technique to these applications we are developing special micro-porous substrates, which consist of high-density individually fillable micro-volumes as depicted in equipment. In this case, the irradiation dose of sample is reduced by more than two orders of magnitude compared to µ-PES setup thus eliminating water radiolysis related restrictions.

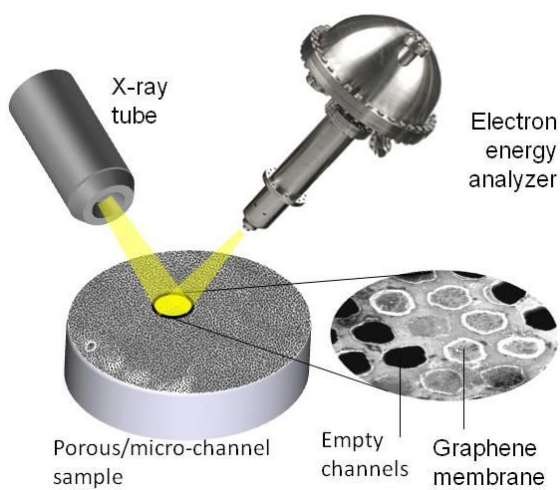

**Figure 6.** The principal concept of micro-porous substrates, which consist of the high-density individual micro-volumes separated from the vacuum by electron transparent membrane. The sample can be filled (impregnated) with the analyte which can be analyzed with standard laboratory XPS equipment.

Fig. 6. On the front side of such a sample, nearly all micro-volumes can be vacuum-sealed by the electron transparent membrane. An incidental disruption of a fraction of the individual micro-windows will not lead to complete sample loss since every micro-volume is isolated from the others. The filling factor of this kind of sample can be as high as 50 %, making these samples suitable for analysis by standard PES

**Conclusions**

In conclusion, we have evaluated the transmissivity of single and multi-layer graphene for photoelectrons with kinetic energies in the range from ≈100 eV to 1000 eV and different geometries of the PES setup. We have demonstrated the experimental approach as well as principal possibility to acquire photoelectron spectra from fully hydrated interfaces using graphene as an electron, X-ray, and optically transparent window separating UHV from the sample environment. Further development of this methodology will allow low-cost ambient pressure PES to overcome some of the current experimental challenges and the requirements for using expensive differentially pumped detection systems. In particular, this alternative approach is very appropriate for studies of liquid or gaseous toxic, highly reactive, or even radioactive samples and interfaces that will be enclosed inside the hermetic environmental cell and analyzed through the small electron-transparent cell window. The class of suitable ultrathin window materials is not limited to graphene and its derivatives only. Recently high yield protocols for the fabrication of 2D and quasi-2D membranes made of BN, $MoS_2$, oxides, and other materials have been reported [33]. This versatility in selection of electron-transparent window material will complement and broaden the possibilities of environmental electron spectroscopy and microscopy. One of the major limitations we found when attempting high spatial resolution studies are X-ray or electron beam-induced effects when critical radiation doses exceed a certain threshold, which has been already extensively reported in high-resolution transmission electron microscopy (HRTEM) and SEM studies. For that reason, for imaging of fully hydrated samples for extended periods of time, one needs to tune the irradiation conditions not to exceed the "radiolysis dose threshold". Alternatively, we propose the design of the sample platform for liquid samples which significantly relaxes aforementioned constrains. Such a platform which can also be used with other spectroscopic techniques sensitive to liquid (gas)-solid interfaces[34] is currently under the tests.

**Methods**

**Graphene samples**. Monolayer graphene (g-Cu) was grown on 25 µm thick Cu foils at a temperature of ≈1000°C in a $CH_4/H_2$ reactive atmosphere within the hPa pressure range following a protocol described in Ref [22]. For the fabrication of more robust graphene windows for liquid cells, membranes with an average thickness of 4 monolayers were CVD grown on Ni films deposited on Si wafers. All samples were transported through air to the synchrotron facility ELETTRA, where further device preparation was performed prior to the µ-PES measurements.

To prepare suspended membranes we have tested two different approaches. In the first method the Cu underneath of the as-grown graphene was electrochemically etched. The etching was conducted while monitoring the top side of the sample with optical microscopy and illumination of the sample backside. The etching was halted as soon as first few micrometer-sized holes appeared. In order to keep the graphene as clean as possible, a special etching procedure was applied, avoiding the placement of any protection layer on top of graphene, since such a layer eventually contaminates the suspended graphene membrane (see supporting information). As a result, such prepared suspended membranes should have the highest electron transparency.

The second type of membranes was made via transferring of the graphene layer on to a Au-coated stainless steel support which contained a few micrometers' wide orifice using a modified PMMA transfer protocol (see details in [35]). To record the spectra from the wet sample through the graphene membrane, we used a custom-made single use E-cell described elsewhere [20a, 20c]. The details of the graphene fabrication, SEM and Raman characterization, as well as the E-cell features, can be found in the supporting information.

**Photoelectron spectroscopy and microscopy**. Photoelectron attenuation tests were performed on as-grown graphene on Cu substrate foils (g-Cu) using a laboratory XPS system equipped with a non-monochromatic Mg Kα and Al Kα photon source. The PE signal intensity increase after the removal of the grown carbon layer by $Ar^+$ ion etching was measured and related to the effective attenuation length of the emitted photoelectrons in graphene. The largest error affecting the precision of the attenuation tests relies in the fact that the XPS apparatus has to be switched off when removing the grown graphene layer by $Ar^+$ ion etching. The systematic error of ± 5 % (as determined in separate tests) when reproducibly adjusting the same X-ray photon flux and electron detection efficiency upon switching on again the X-ray system significantly exceeds all statistical errors, which can be kept low by sufficiently increasing the detection time. Thus, this systematic error affects the extracted effective attenuation lengths which are indicated as error bars in Fig. 1 and which are calculated by assuming the error propagation according to:

$$\Delta \lambda_{EAL} \approx \frac{d\lambda_{EAL}}{d(I_0/I)}\Delta(I_0/I) = -\frac{d_G}{\cos\theta}\frac{I}{I_0}\left(\frac{1}{\ln(I_0/I)}\right)^2 \Delta(I_0/I)$$

For electron attenuation tests in the μ-PES setup with as grown (g-Cu) and wet-transferred graphene samples (g-Au), XPS substrate spectra were collected comparatively from the graphene-covered and adjacent pristine areas of the support surface (Fig.2 and 3). For this purpose we prepared two different kinds of samples: (i) part of the as-grown g-Cu samples was sputtered through a shadow mask thus exposing a "pristine" Cu area (Fig. 2a-c); (ii) for the transferred graphene samples the graphene edge was located and signals from uncovered and g-covered Au areas were compared (Fig. 2d).

The μ-PES measurements were performed using the scanning photoelectron microscope (SPEM) at the ELETTRA ESCA-microscopy beamline [36]. In the SPEM setup, Fresnel zone plate optics was used to focus the X-ray beam onto a spot of *ca.* 100 nm in diameter (Fig. 2e). The chemical, topography and transmission mapping of the sample can be obtained via raster scanning of the sample with respect to the focused X-ray beam with simultaneous collection of the element specific photoelectrons or transmitted photons. Depending on the raster step, micrometer or millimeter-sized chemical maps can be recorded. The maximum spatial resolution amounts ≈ 0.1 μm. Detailed PES spectra can be acquired at any specific location selected from the maps. Depending on the substrate, Au 4f and Cu 2p, Cu 3p and C 1s core levels were selected in the present study for the spectroscopic and signal attenuation tests since they have advantageous combination of appreciable photo ionization cross section (from about 0.1 Mb to 1 Mb) for hv ≈ $10^3$ eV used in the present study and relatively low binding energy. The latter offers high enough kinetic energy of the Au 4f and Cu 2p photoelectrons for penetration through the graphene layers. The incident angle of the X-ray beam and emission angle $\Theta$ of the electron analyzer were kept at 0° and 60°, respectively, with respect to the sample normal unless specified differently (see Fig. 2e and Fig. 3a). For quantitative analysis of the acquired data, the photoemission peaks were fitted by Doniach-Sunjic line shapes[37].

**Liquid sample preparation.** Ultrapure 10 µL water droplet sample was pipetted on to the back side of the Au coated stainless steel disc with graphene covered micro-orifice. Inside E-cell assembly the sample was sealed with elastomer membrane. After the sealing, the E-cell was loaded at first in an intermediate load lock chamber, which was gradually evacuated, and then transferred to UHV SPEM chamber. XPS imaging and spectroscopic measurements were performed *ca.* 1 hour after the cell was first exposed to vacuum. After completion of the experimental run, the E-cell was inspected to confirm the presence of liquid water inside the cell. The observed electron beam-induced dynamics under the membrane (such as bubble formation, see main text) was indicative that the graphene membrane preserved its integrity during the µ-PES measurements and that the liquid water was present inside the E-cell during the measurements. To minimize X-ray beam-induced damage of the g-membrane, care was taken to reduce the irradiation dose during the spectra acquisition by adjusting the X-ray focus accordingly. To avoid unnecessary exposure during the "parking" of the X-ray beam, the centered light spot was offset with respect to the center of the g-membrane resulting in a beam parking position outside the membrane.

*Conflict of Interest:* The authors declare no competing financial interest.

*Acknowledgements:* The authors would like to thank Mr. Joshua Stoll and Clay Watts (SIUC) for their help in preparation of the experiment. The suggestions and comments from Drs. N. Zhitenev, Ch. Brown, D. Meier, R. Sharma and C. Powell (all at NIST) are greatly acknowledged. This work made extensive use NIST SRD-82 EAL and NIST -71 IMFP databases, for which the authors are thankful to Drs. C. J. Powell and A. Jablonski. S.G. gratefully acknowledges financial support from the German Science Foundation (DFG, contract number GU 521/2-1).*Supporting information available:* Graphene grown and handling protocols as well as characterization procedures. Preparation protocol for the local formation of suspended graphene on Cu. Relation of the topographic contrast in the photoemission images to the local surface geometry and estimation of the achievable photoelectron yield. SEM movie of the bubble formation under graphene membrane as a result of electron beam irradiation of water.